\documentclass[10pt,aps,prl,twocolumn,superscriptaddress,floatfix,longbibliography]{revtex4-1}

\usepackage{graphicx}
\usepackage{amsfonts}
\usepackage{amssymb}
\usepackage{amsmath}
\usepackage{txfonts}
\usepackage{lipsum}
\usepackage[dvipsnames]{xcolor}
\usepackage{wasysym}
\usepackage[colorlinks=true, allcolors=blue]{hyperref}
\usepackage{bbold}
\usepackage{etoolbox} 
\usepackage[normalem]{ulem}
\usepackage{mathtools} 
\usepackage{multirow} 
\usepackage{hhline}
\usepackage{mathrsfs} 
\usepackage{soul}
\usepackage{array}

\usepackage[inline]{enumitem}

\usepackage{letltxmacro}
\LetLtxMacro{\oldsqrt}{\sqrt}
\renewcommand{\sqrt}[2][\mkern8mu]{\mkern-6mu\mathop{}\oldsqrt[#1]{#2}}

\newcommand{\SRO}{Sr$_2$RuO$_4$\,}

\begin{document}

\title{From a Lifshitz Transition to Spin-Density-Wave Order: Sr$_2$RuO$_4$ under Uniaxial Strain}

\author{Evgeny A. Stepanov}
\affiliation{CPHT, CNRS, {\'E}cole polytechnique, Institut Polytechnique de Paris, 91120 Palaiseau, France}
\affiliation{Coll{\`e}ge de France, Universit{\'e} PSL, 11 place Marcelin Berthelot, 75005 Paris, France}

\author{Sophie Beck}
\affiliation{Institute of Solid State Physics, TU Wien, 1040 Vienna, Austria}

\author{Maria Chatzieleftheriou}
\affiliation{CPHT, CNRS, {\'E}cole polytechnique, Institut Polytechnique de Paris, 91120 Palaiseau, France}

\author{Antoine Georges}
\affiliation{Center for Computational Quantum Physics, Flatiron Institute, New York 10010, USA}
\affiliation{Coll{\`e}ge de France, Universit{\'e} PSL, 11 place Marcelin Berthelot, 75005 Paris, France}
\affiliation{CPHT, CNRS, {\'E}cole polytechnique, Institut Polytechnique de Paris, 91120 Palaiseau, France}
\affiliation{DQMP, Universit{\'e} de Gen{\`e}ve, 24 quai Ernest Ansermet, CH-1211 Gen{\`e}ve, Suisse}

\begin{abstract}
In recent years, the effects of uniaxial strain on Sr$_2$RuO$_4$ have been extensively investigated, revealing a pronounced enhancement of the superconducting critical temperature accompanied by experimental indications of an emergent magnetically ordered state. 
In this Letter, we study the impact of strain on the spectral and magnetic properties of this material within an advanced many-body theoretical framework.
We find that strain induces an anisotropic Lifshitz transition, which enhances the electron scattering rate along the elongation axis while simultaneously promoting coherent Fermi-liquid behavior in the perpendicular direction, in close agreement with experimental observations.
Taking into account this anisotropic behavior allows us to capture the magnetic instability leading to a finite-temperature ordered state at large strain values.
Our analysis further reveals that the magnetic response is dominated by two competing wave vectors associated with distinct nesting vectors: one within the $\gamma$ sheet and the other between the $\alpha$ and $\beta$ sheets of the Fermi surface. 
Both wave vectors can stabilize magnetic order, with their competition depending sensitively on the coherence of the respective Fermi surface sheets.
\end{abstract}

\maketitle

The ruthenate compound Sr$_2$RuO$_4$ has become a paradigmatic correlated electron system exhibiting a variety of remarkable many-body effects~\cite{PhysRevLett.123.247001, PhysRevResearch.4.033011}. 
Its realization in clean, high-quality single crystals has enabled extensive experimental investigation. 
In the normal state, the material displays a strong electronic mass enhancement and pronounced spin fluctuations, yet it does not develop magnetic order at low temperatures~\cite{PhysRevLett.76.3786, PhysRevLett.85.5194, PhysRevB.72.205114}. 
Instead, Sr$_2$RuO$_4$ enters a superconducting state around 1 K~\cite{maeno1994superconductivity}.
More recently, experiments have shown that the superconducting transition temperature can be significantly enhanced under uniaxial strain~\cite{doi:10.1126/science.1248292, doi:10.1126/science.aaf9398, doi:10.1073/pnas.2020492118, doi:10.1126/science.adf3348}. 

There is broad consensus that superconductivity in Sr$_2$RuO$_4$ is closely linked to strong magnetic fluctuations~\cite{RevModPhys.75.657, doi:10.1143/JPSJ.81.011009,profe2024competition, doi:10.1073/pnas.2025313118, PhysRevB.108.094516, PhysRevLett.97.167002, luke1998time, doi:10.7566/JPSJ.93.062001, benhabib2021ultrasound}, making their accurate characterization of central importance.
Inelastic neutron scattering experiments on unstrained Sr$_2$RuO$_4$ reveal a dominant magnetic response at the wave vector ${Q=(3\pi/5,3\pi/5,0)}$, associated with nesting between the $\alpha$ and $\beta$ Fermi surface (FS) sheets, together with a broader response centered at the Brillouin zone (BZ) center~\cite{PhysRevLett.83.3320, PhysRevB.66.064522, PhysRevLett.92.097402, PhysRevLett.122.047004, PhysRevB.103.104511}.
In contrast, much less is known about the magnetic response under applied strain. 
Recent measurements based on muon-spin relaxation, the elastocaloric effect, and Hall resistivity provide indirect evidence for the emergence of a spin-density wave (SDW) order at sufficiently large strain~\cite{grinenko2021split, li2022elastocaloric, PhysRevLett.131.036301}. 
While the available data suggest that the ordered state is incommensurate, its ordering wave vector and the form of the magnetic response remain unknown.

Understanding the evolution of Sr$_2$RuO$_4$ under uniaxial strain remains a major theoretical challenge. The application of strain drives a Lifshitz transition of the $\gamma$ FS sheet, whose topology changes as a van Hove singularity crosses the Fermi level. 
This transition gives rise to a strong anisotropy, with pronounced non-Fermi-liquid behavior along the direction of strain and a clear Fermi-liquid (FL) character in the perpendicular direction, as recently revealed in an angular-resolved photoemission (ARPES) experiment~\cite{hunter2025non}. 
The presence of strong electronic correlations necessitates the use of non-perturbative methods, such as dynamical mean-field theory (DMFT)~\cite{RevModPhys.68.13}, which go beyond the weak-coupling limit. 
DMFT calculations reproduce the FS topology in the unstrained case in good agreement with experiment~\cite{PhysRevX.9.021048, PhysRevLett.84.2662, PhysRevLett.85.5194, PhysRevB.74.035115}, as well as many physical properties of \SRO in its normal state~\cite{PhysRevLett.106.096401}. 
However, the strongly anisotropic FL {\it vs} non-FL behavior along the FS under strain cannot be captured by single-site DMFT, which replaces the momentum-dependent electronic self-energy by its average over the BZ.

The magnetic response of Sr$_2$RuO$_4$ is even more challenging to address theoretically already in the unstrained case.
Single-site DMFT was shown to overcome several shortcomings of the RPA 
approximation~\cite{PhysRevB.100.125120}, but it was shown in Ref.~\cite{ts6y-zb6m} 
that an accurate description must include both strong local correlations and spatial collective electronic fluctuations. 
The evolution of the magnetic signal under strain is also not well understood, neither experimentally nor theoretically. 
Random-phase approximation (RPA) calculations predict that strain suppresses magnetic fluctuations at the original $Q$ vector and induces new fluctuations at ${Q'=(\pm\pi, \pm\pi/2)}$, associated with secondary FS nesting of the deformed $\gamma$ sheet emerging beyond the Lifshitz transition~\cite{PhysRevB.102.054506}. 
However, this weak-coupling approach cannot account for incoherent behavior and correlation-induced renormalization effects~\cite{ts6y-zb6m}, which may strongly affect magnetic fluctuations. 
More advanced functional renormalization group calculations do not capture the strain-induced SDW state found in RPA~\cite{profe2024competition}. 
These discrepancies call for a theoretical framework beyond standard weak-coupling approaches.

\begin{figure}[t!]
\centering
\includegraphics[width=\linewidth]{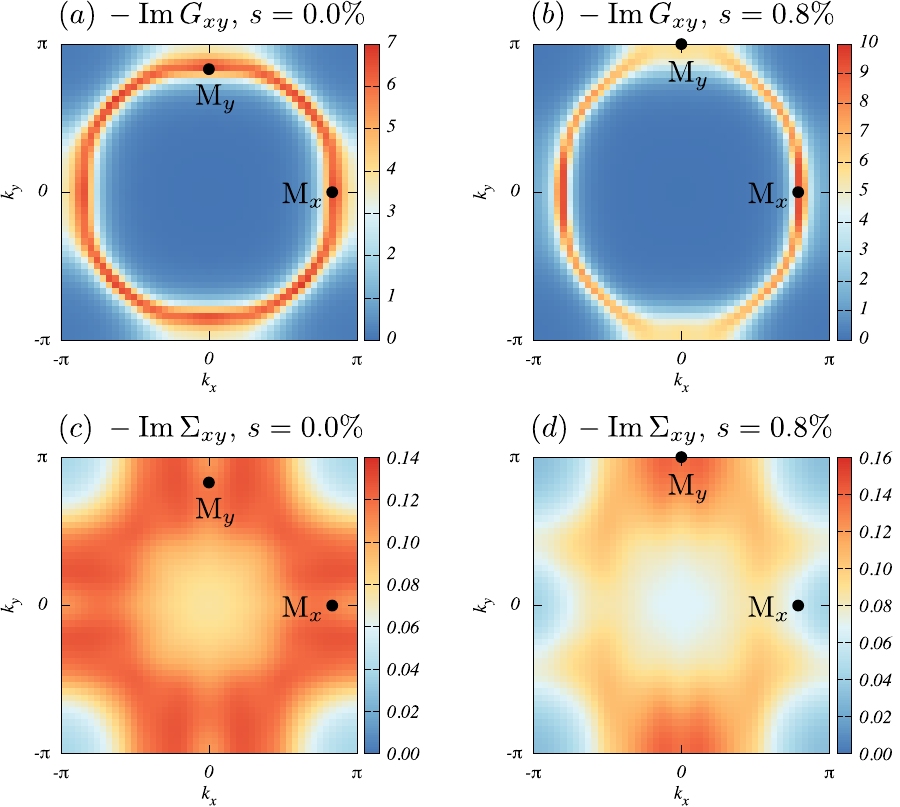}
\caption{{\it Momentum differentiation under strain.} 
Top row: Proxy [$-\mathrm{Im}G_{xy}(\nu_0)$] for the spectral density at the Fermi energy of the $xy$ orbital 
at ${T=145}$\,K for (a) ${s=0\%}$ and (b) ${s=0.8\%}$ strains. 
Bottom row: Proxy for the scattering rate of the $xy$ orbital [$-\mathrm{Im}\Sigma_{xy}(\nu_0)$] for (c) ${s=0\%}$ and (d) ${s=0.8\%}$ strains. 
The black dots depict the M$_x$ and M$_y$ points on the FS.
\label{fig:S_BZ}}
\end{figure}

In this Letter, we investigate Sr$_2$RuO$_4$ within an advanced many-body theoretical framework that consistently accounts for all key ingredients necessary to capture the evolution of electronic correlations and magnetic fluctuations under uniaxial strain. 
We show that the strain-induced deformation of the FS leads to a strongly momentum-dependent anisotropic self-energy and a pronounced enhancement of the quasiparticle scattering rate at the Lifshitz transition, in agreement with experiment~\cite{hunter2025non}. 
We further find that increasing strain promotes magnetic fluctuations at the $Q'$ wave vector, associated with secondary nesting of the deformed $\gamma$ sheet of the FS. 
However, the highly incoherent behavior of this sheet, inherited from the Lifshitz transition, persists up to larger strain values and suppresses fluctuations at $Q'$, placing them in strong competition with the SDW instability at a wave vector $Q$ close to the original one.
At large strain, SDW order can emerge at either wave vector, with the dominant ordering strongly dependent on the coherence of the corresponding FS sheet.

\begin{figure}[h!]
\centering
\includegraphics[width=1\linewidth]{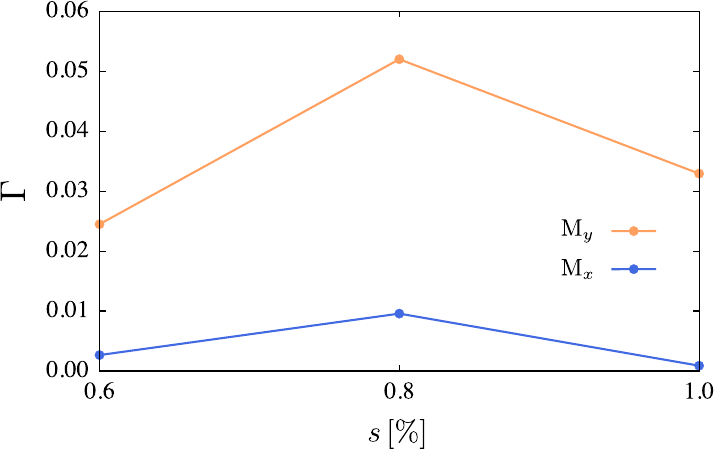}
\caption{\textit{Quasiparticle scattering rate of the $xy$ orbital.} $\Gamma$ is plotted as a function of strain, around the Lifshitz transition, calculated at the M$_y$ (orange) and M$_x$ (blue) points on the FS, at ${T=145}$\,K.
The non-monotonic behavior of the scattering rate at M$_y$ upon increasing strain reveals that the Lifshitz transition takes place at ${s\simeq0.8\%}$.
\label{fig:S_v}}
\end{figure}

Although local correlations as described for by single-site DMFT account for 
many properties of \SRO~\cite{PhysRevLett.106.096401}, the importance of non-local effects 
has become evident in recent years~\cite{ts6y-zb6m,PhysRevLett.116.256401, acharya2019evening}. 
In this work, we utilize the standard {\it ab-initio} derived three-orbital model for Sr$_2$RuO$_4$, discussed in detail in the End Matter.
The model is solved within an advanced many-body method, the dual triply irreducible local expansion ({\mbox{D-TRILEX}) approach~\cite{PhysRevB.100.205115, PhysRevB.103.245123, 10.21468/SciPostPhys.13.2.036}.
This non-perturbative method incorporates orbital-dependent charge and spin local and non-local fluctuations diagrammatically beyond DMFT~\cite{RevModPhys.68.13}. 
It enables a self-consistent treatment of the interplay between collective electronic fluctuations and single-particle quantities~\cite{stepanov2021coexisting, Vandelli2024_PbSi, PhysRevB.110.L161106, PhysRevLett.132.236504, j6bj-gz7j, Cuprates, 46h3-sprg}. 
At the same time, the tractable diagrammatic structure of the method enables efficient yet accurate calculations in multi-orbital~\cite{PhysRevLett.127.207205, PhysRevLett.129.096404, PhysRevResearch.5.L022016, PhysRevLett.132.226501, stepanov2024charge, ts6y-zb6m, dhm8-5ss6, c9ch-w4vk, cluster_topology} and real-time~\cite{vglv-2rmv, gzwp-zd6t, N-AN-dynamics} cases.
Importantly, this approach enabled an accurate description of both spectral properties and magnetic response of the unstrained system in agreement with experiments~\cite{ts6y-zb6m}.

\begin{figure}[t!]
\centering
\includegraphics[width=\linewidth]{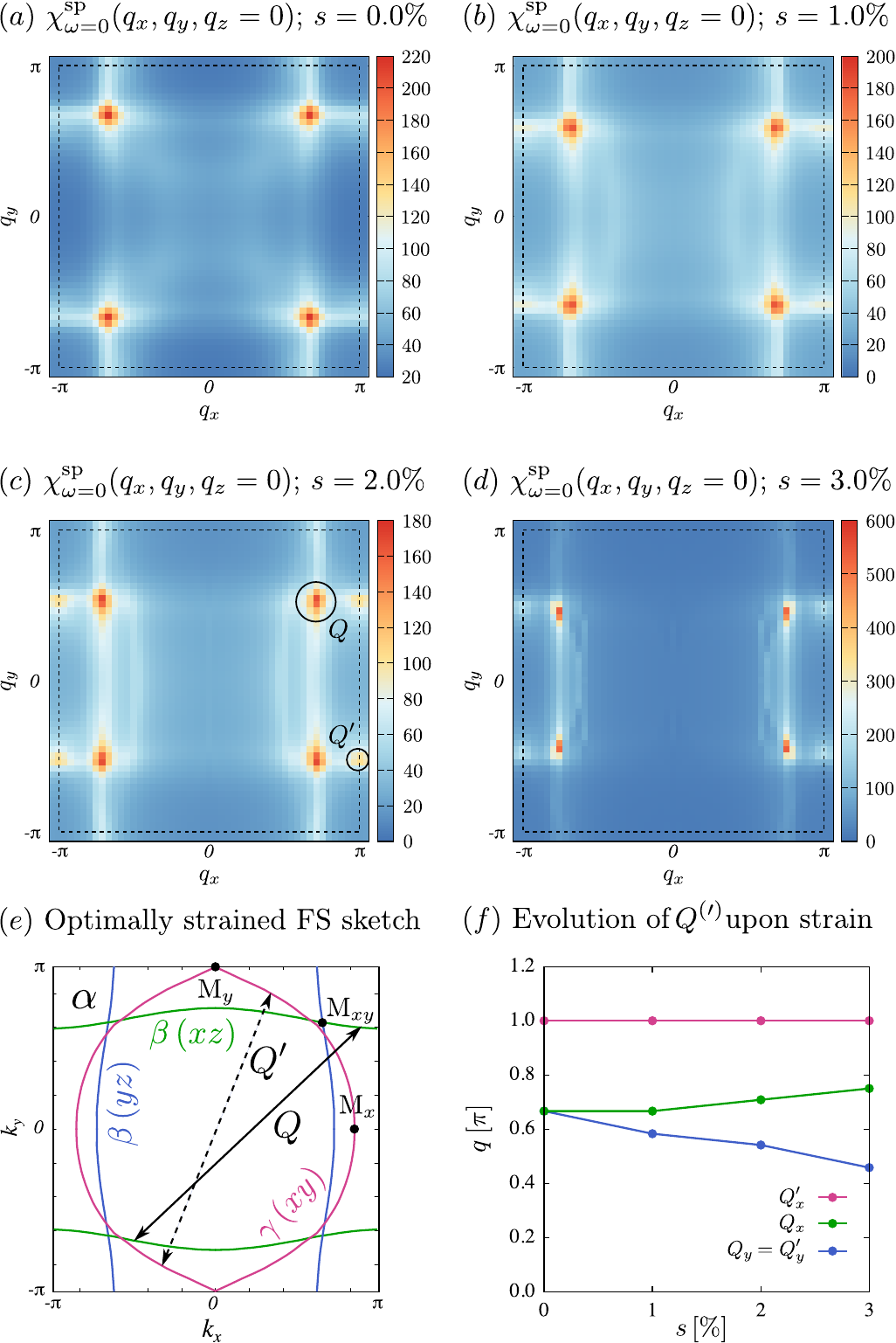}
\caption{
{\it Momentum dependence of the magnetic fluctuations as a function of strain.}
Panels (a) to (d): Static spin susceptibility calculated on the $(q_x, q_y, 0)$ plane for $0.0\%$, $1.0\%$, $2.0\%$, 
and $3.0\%$ of strain for temperature ${T=145}$\,K. The BZ boundary is depicted by the dashed black square.
At high strains a subleading intensity is developed at the $Q'$ point, linked to the nesting vector shown schematically in (e). The position of the $Q$ and $Q'$ on panel (c) is indicated by the black circles.
Upon increasing strain, the FS deformation modifies the momentum-dependence of the spin susceptibility, bringing the $q_y$ position of $Q$ and $Q'$ wave vectors closer to the center of the BZ, as depicted in (f), shifting $Q_{x}$ to slightly larger values of $q$, while the $Q'_x$ position remain nearly unchanged.
\label{fig:X_BZ}}
\end{figure}

Experimental studies indicate that strain primarily affects the two-dimensional $d_{xy}$ orbital and that the associated $\gamma$ sheet of the FS undergoes the Lifshitz transition discussed above at a strain of approximately $0.6\%$~\cite{steppke2017strong, sunko2019direct}.
In density functional theory calculations, this transition is found at larger values of strain ($>1.1\%$)~\cite{PhysRevB.94.045118, PhysRevB.100.245139}, while other studies have retrieved the experimental value upon inclusion of a large spin-orbit coupling (SOC)~\cite{steppke2017strong, PhysRevLett.130.096401}. 
The \mbox{D-TRILEX} simulations, performed without SOC, find the Lifshitz transition at $0.8\%$ of strain, in reasonable agreement with experiments and previous DMFT calculations~\cite{profe2024competition}. 
Although the inclusion of SOC is important for accurately describing the details of FS at low temperatures, its effect on the spin susceptibility is expected to be less significant at the relatively high temperatures considered in this work~\cite{PhysRevB.65.220502, PhysRevLett.92.097402}. 
Incorporating SOC into advanced many-body approaches capable of capturing strong electronic correlations remains a significant theoretical challenge. 
We therefore expect our results to reliably capture the qualitative evolution of the electronic and magnetic properties under strain, while the predicted magnetic ordering temperatures should be regarded as qualitative estimates.

As shown on Figure~\ref{fig:S_BZ}, we find that the system displays 
an increasing anisotropy along the FS as 
the Lifshitz transition and corresponding van Hove singularity is approached under uniaxial strain. 
The figure displays the negative of the Green's function (a,\,b) and of the self-energy (c,\,d) of the $d_{xy}$ orbital at the lowest Matsubara frequency ($\nu_0$) in the BZ for ${k_z=0}$, which are proxies for the spectral function and the scattering rate, respectively.
The largest intensity contour in panels (a) and (b) therefore visualizes the $\gamma$ sheet of the FS.
At zero strain, the self-energy along the $\gamma$ sheet is nearly momentum-independent (c), consistent with previous calculations~\cite{ts6y-zb6m}. 
Applying $0.8\%$ of strain along the $y$ direction induces a pronounced anisotropic momentum dependence of the self-energy (d).
In particular, the imaginary part of the self-energy (scattering rate) at the M$_y$ point, located along the elongation axis, becomes substantially larger than at the M$_x$ point on the perpendicular axis. 

The momentum anisotropy of the self-energy indicates a pronounced loss of coherence at the M$_y$ point, associated with the Lifshitz transition. 
In Figure~\ref{fig:S_v}, the scattering rate $\Gamma$ of the $d_{xy}$ orbital is plotted as a function of strain in the vicinity of the Lifshitz transition (${s=0.8\%}$), at the M$_x$ and M$_y$ points depicted in Figure~\ref{fig:S_BZ}, calculated as the product of ${\text{Im}\,\Sigma_{xy}(\nu\to0)}$ and the quasiparticle residue ${\text{Z}=\big(1-\frac{\partial \text{Im}\Sigma}{\partial\nu}\big|_{\nu\rightarrow0}\big)^{-1}}$: ${\Gamma=\text{Im}\,\Sigma_{xy}(\nu\to0)\cdot \text{Z}}$, obtained upon Matsubara data fitting. 
For ${s=1.0\%}$, beyond the Lifshitz transition, the FS changes its topology, and the scattering rate is plotted at the FS ${\bf k}$ point closest to M$_y$.
We find that the scattering rate at the M$_x$ point is very small, exhibiting clear FL behavior.
In contrast, the M$_y$ point 
displays much more correlated behavior, with a finite scattering rate which increases approximately two times at $0.8\%$ of strain, compared to both $0.6\%$ and $1.0\%$. 
These results are in excellent agreement with recent experiments, which observe a peak in the quasiparticle scattering rate at the Lifshitz transition, reaching approximately twice the value at smaller strain~\cite{hunter2025non}.
As we show further, this strain-induced anisotropic incoherent behavior plays an important role in the formation of the SDW ordered state.

Let us now investigate the evolution of magnetic fluctuations under applied strain. 
Figure~\ref{fig:X_BZ} shows the static (${\omega=0}$) spin susceptibility $\chi_{\rm sp}({\bf q},\omega) = \sum_{l,l'} \chi^{\rm sp}_{ll'}({\bf q},\omega) = \langle m_{{\bf q},\omega,l} \, m_{{\bf -q},-\omega,l'} \rangle$, where ${m_{l} = n_{l\uparrow} - n_{l\downarrow}}$ is the spin density in orbital ${l\in\{xz,yz,xy\}}$.
The susceptibility is plotted over the first BZ (black dashed square) for increasing strain (from (a) to (d)) at ${T=145}$\,K.
In the unstrained case, the leading magnetic fluctuations occur at the incommensurate wave vector ${Q=(3\pi/5,3\pi/5,0)}$, as evidenced by the Bragg peaks in panel (a) of Figure~\ref{fig:X_BZ}. 
This signal originates from intra-orbital nesting of the $\alpha$ and $\beta$ FS sheets of the $d_{xz}$ and $d_{yz}$ orbitals, shown schematically in panel (e).
The application of strain gives rise to two prominent effects. First, the strain-induced deformation of the $\gamma$ FS sheet beyond the Lifshitz transition gives rise to a subleading magnetic signal at ${Q'\simeq(\pi,\pi/2,0)}$~\cite{PhysRevLett.123.247001, PhysRevB.102.054506, profe2024competition}. 
This wave vector, associated with intra-orbital $d_{xy}$ nesting, is indicated by the dashed arrow in panel (e). The corresponding peak in the spin susceptibility is already visible at $1.0\%$ strain (panel (b)) and becomes increasingly pronounced at $2.0\%$ strain.
Second, the FS distortion shifts both $Q$ and $Q'$ toward smaller values of $q_y$ and only slightly affects the $Q_x$ position while preserving ${Q'_{x}=\pi}$, as illustrated in panel (f). However, at the temperature considered, strain does not change the relative strength of the two magnetic fluctuations: the $Q$ mode, already present in the unstrained system, remains dominant. 
At $3.0\%$ strain, the spin susceptibility develops a pronounced peak at $Q$ (panel (d)), signaling a strong tendency toward SDW order.
Nevertheless, the spin susceptibility at $Q'$ remains substantial, indicating that the corresponding magnetic fluctuations continue to play an important role.

\begin{figure}[t!]
\includegraphics[width=1\linewidth]{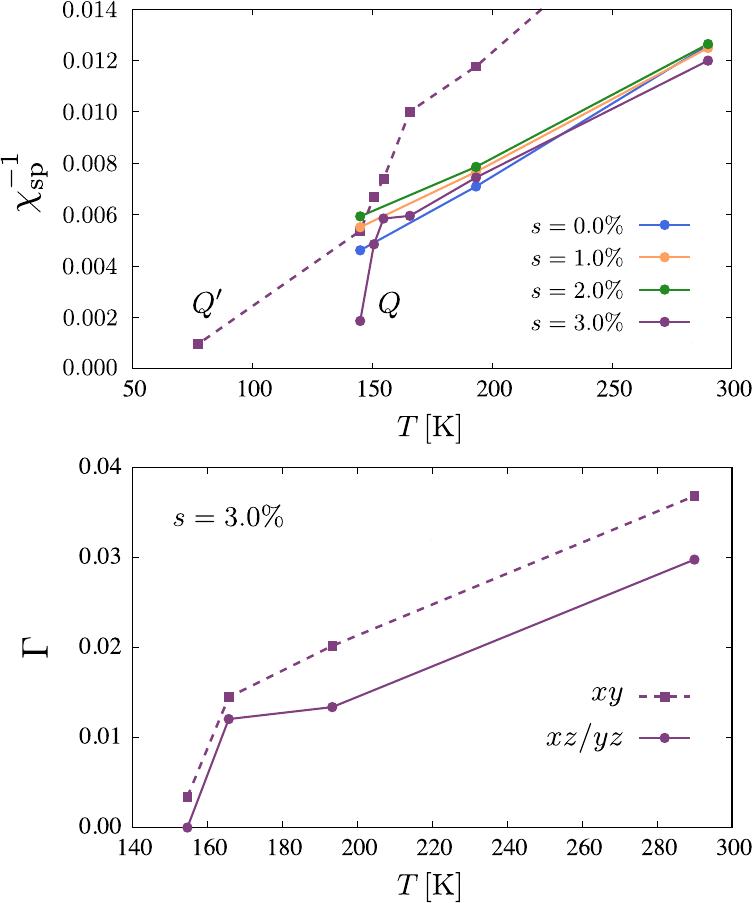}
\caption{\textit{Inverse of the spin susceptibility and scattering rate as a function of temperature.} Top panel: In solid lines, $\chi_{sp}^{-1}$ is calculated at the $Q$ wave vector where it is largest, as a function of temperature for $0.0\%$, $1.0\%$, $2.0\%$, and $3.0\%$ of strain. At $3.0\%$, $\chi_{sp}^{-1}$ is shown at the $Q'$ vector with a dashed line, revealing the strong competition between the two ordering vectors at low T.
Bottom panel: The scattering rate of the $xy$ orbital as a function of temperature, calculated for $3.0\%$ of strain at the M$_y$ point for the $xy$ orbital and at the corresponding $q-$point for the $xz/yz$ orbitals. The crossover from a bad-metal to a good FL for both vectors is revealed at low T.
\label{fig:X_T}}
\end{figure}

To analyze the formation of the SDW state, in the upper panel of Figure~\ref{fig:X_T} we plot the inverse of the leading spin susceptibility as a function of temperature for different values of strain. 
At $0.0\%$, $1.0\%$, and $2.0\%$ of strain a linear dependence on $T$ is observed, with a slope that implies the absence of an ordered state at low temperatures. 
The increase to a larger strain value of $3.0\%$ induces a qualitative change in the behavior of the susceptibility, which appears to be split into two temperature regimes. 
In particular, for $3.0\%$ we compare the susceptibility at the $Q$ vector (solid magenta line) and at the $Q'$ one (dashed magenta line).
At high temperatures, ${T>160}$\,K, the linear trend of smaller strain values is preserved for both vectors, with $\chi_{sp}(Q')$ exhibiting significantly smaller values compared to the leading $\chi_{sp}(Q)$. 
At lower temperatures, the two susceptibility curves exhibit an abrupt change in slope, signaling the onset of magnetic order at a finite temperature of ${T\simeq140}$\,K, as indicated by the extrapolated curve. 
We emphasize that this should not be taken as a quantitative estimate of the ordering temperature but 
rather as an upper bound: additional fluctuations not included here or the omitted SOC will lower it further, as indeed suggested by experiments\cite{PhysRevLett.131.036301}.
Remarkably, below ${T\simeq170}$\,K the susceptibility at the two vectors is very comparable, signaling a tight competition between two ordering tendencies. In fact, we were able to stabilize a calculation at the very low ${T=77}$\,K, performed with a slightly larger chemical-potential shift (${\Delta\mu=0.13}$\,eV), revealing that, for essentially the same electronic filling, a SDW state with ordering vector $Q'$ can also be stabilized.
These findings indicate that the system lies close to a competition between the two SDW instabilities, such that a small redistribution of the electronic density between different orbitals can tip the balance in favor of either the $Q$ or the $Q'$ ordered state.

In the lower panel of Figure~\ref{fig:X_T} we plot the scattering rate of the $d_{xy}$ and $d_{xz/yz}$ orbitals as a function of temperature for $3.0\%$ of strain, at the M$_y$ and M$_{xy}$ points, respectively (see Fig.~\ref{fig:X_BZ}\,(e)), where the scattering rate is maximal. 
At high temperature, the scattering rate of $d_{xy}$ is consistently larger than the one of $d_{xz/yz}$, with both of them revealing a pronounced incoherent behavior, although stronger for $d_{xy}$.
This incoherence, although reduced, remains substantial down to ${T\approx170}$\,K, thereby suppressing the magnetic fluctuations of primarily the $Q'$ vector arising from nesting of the corresponding FS sheet.
However, by further lowering the temperature the system progressively retrieves its lost coherence, gradually crossing over to a good FL state.
Below ${T\approx170}$\,K, the scattering rate becomes very small (close to zero) and comparable for both orbitals, which is linked with the low temperature regime of spin susceptibility, characterized by a strong competition between the two ordering tendencies. 

The corresponding behavior of the single-particle (scattering rate) and the two-particle (susceptibility) quantities in two distinct temperature regimes reveals that the increase of coherence upon lowering $T$ leads to a better FS nesting, promoting larger magnetic fluctuations, eventually culminating in an ordered state. 
In the End Matter a discussion on the orbital dependence of the spin susceptibility is presented. 
It is illustrated that at high temperatures the magnetic response is heavily broadened. At lower temperatures the signal becomes amplified and progressively sharper, especially at the $Q'$ vector due to the better FS nesting of the $\gamma$ sheet.
Nevertheless, it still cannot fully destabilize the SDW ordering associated with the $Q$ vector.
This result reveals the competing nature of the magnetic response at two vectors with predominantly different orbital character.

To conclude, we have investigated the evolution of nonlocal correlation effects and magnetic response in Sr$_2$RuO$_4$ under the effect of uniaxial strain. 
We find that, upon approaching the Lifshitz transition of the $\gamma$ Fermi-surface sheet, the self-energy of the corresponding $d_{xy}$ orbital develops a substantial momentum dependence. Along the elongation axis, the self-energy acquires a bad metallic/highly incoherent behavior, characterized by an increased scattering rate by approximately a factor of two, compared to the unstrained case, in agreement with the experiment~\cite{hunter2025non}.
In turn, the self-energy in the perpendicular direction displays a very good FL behavior with a nearly zero scattering rate.  

We have observed that the deformation of the FS by strain affects the form of the spin susceptibility by inducing a subleading mode at the $Q'$ vector, associated with secondary nesting of the $\gamma$ sheet of the Fermi surface, and by shifting both the leading $Q$ and subleading $Q'$ wave vectors toward smaller values of $q_y$} 
We demonstrate that an accurate description of the magnetic response cannot be solely based on the nesting and necessarily requires additional consideration of the anisotropic incoherent behavior induced by electronic correlations beyond the Lifshitz transition.
In contrast to previous weak-coupling RPA calculations, which predict a strong suppression of the $Q$ mode in favor of a single $Q'$ fluctuation upon strain~\cite{PhysRevB.102.054506}, the many-body effects promote 
a strong competition between a SDW state at the $Q$ and the $Q'$ ordering vectors.
This is directly linked to a crossover, below ${T\simeq170}$\,K, into a more coherent regime along most of the FS.
The sharp change in the susceptibility as a function of temperature suggests that the SDW state may persist down to lower strain values at sufficiently low temperatures, which are inaccessible for our calculations.
Our results reveal that Sr$_2$RuO$_4$ forms a SDW ordered state upon uniaxial strain, but they cannot offer a definite answer on the quantitative estimate for the transition temperature and strain value. Regarding the ordering vector we believe that the competition is so tight that very small modifications of the crystal structure can drive the system to either incommensurate SDWs.
Since magnetic fluctuations in Sr$_2$RuO$_4$ are generally believed to be closely linked to superconductivity, the presence of multi-mode spin fluctuations in this system may have important implications for the form of the superconducting order parameter, which remains strongly debated to this day~\cite{PhysRevB.105.155101, profe2024competition}.

\begin{acknowledgments}
The authors thank Felix Baumberger, Stuart Brown, Clifford Hicks, Andrew Hunter, Fabian Kugler, Jonas Profe and Anna Tamai for fruitful discussions.
E.A.S. and M.C. acknowledge support from TGCC-GENCI through the A0180901393 project and also acknowledge EuroHPC Joint Undertaking for awarding access to MareNostrum5 at Barcelona Supercomputing Center (BSC), Spain. 
E.A.S. acknowledges support from ANR JCJC grant ``ELECTRO'', ANR-25-CE30-7064.
The Flatiron Institute is a division of the Simons Foundation.
\end{acknowledgments}

\bibliography{Ref}

\appendix
\setcounter{figure}{0}
\renewcommand{\thefigure}{A\arabic{figure}}
\section{End Matter}
\noindent
\textit{Model --} The strain-dependent electronic structure was obtained from density functional theory calculations performed with the Quantum~ESPRESSO package~\cite{Giannozzi_et_al:2009} using the Perdew-Burke-Ernzerhof (PBE) exchange-correlation functional~\cite{Perdew/Burke/Ernzerhof:1996} and ultrasoft pseudopotentials from the GBRV library~\cite{Garrity_et_al:2014}.
Calculations were carried out using the conventional unit cell sampled with a $10\times10\times3$ $k$-point mesh.

\begin{figure*}[t!]
\centering
\includegraphics[width=\linewidth]{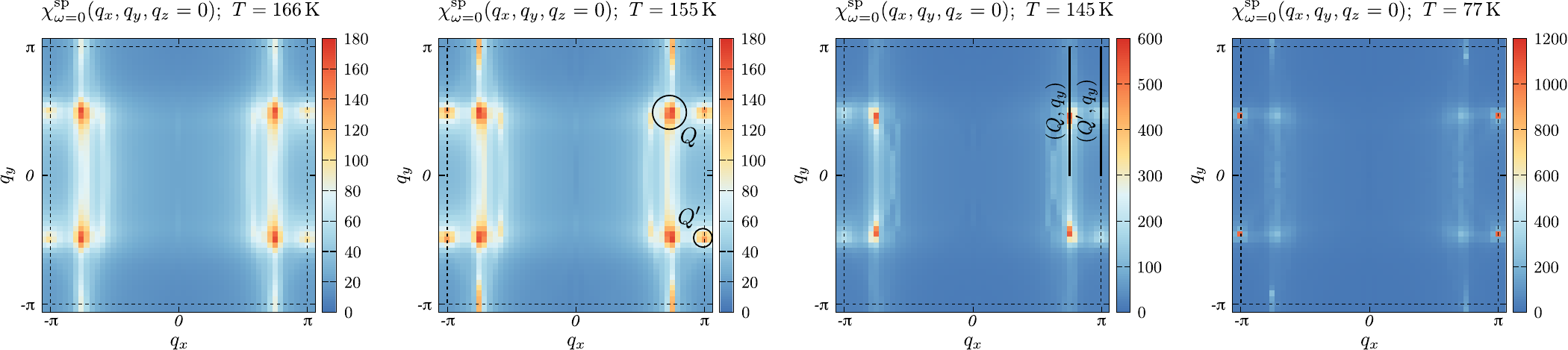}
\caption{\textit{Static spin susceptibility.} Calculations are performed on the $(k_x, k_y, 0)$ plane for 3\% of strain and for temperatures ${T=166}$\,K, ${T=155}$\,K, ${T=145}$\,K, ${T=77}$\,K (from left to right). 
\label{fig:X_BZ_vs_T}}
\end{figure*}

Uniaxial compressive strain was applied along the crystallographic $x$ direction.
Strain values between $0$ and $1.3\%$ were considered in steps of $0.1\%$, together with additional calculations at $2.0\%$ and $3.0\%$ strain.
For each strain value, the lattice parameters perpendicular to the strain direction as well as the internal atomic coordinates were fully relaxed, while maintaining the imposed strain.
Structural relaxations were performed using a pressure convergence threshold of $0.1$ kbar. Spin-orbit coupling was neglected throughout.

The resulting electronic structures were Wannierized using the Wannier90 code~\cite{Pizzi_2020} and mapped onto a low-energy Hamiltonian in the basis of Ru $t_{2g}$ orbitals ($d_{xz}$, $d_{yz}$, and $d_{xy}$). The Wannier Hamiltonians were subsequently transformed to the primitive unit cell and used as input for the many-body calculations.  

In order to account for the many-body effects in Sr$_2$RuO$_4$, we supplement the {\it ab-initio} derived three-orbital model with the local Coulomb interaction parametrized in the Kanamori form~\cite{10.1143/PTP.30.275} with the intraorbital $U$ and interorbital ${U'=U-2J}$ Coulomb interactions, and the Hund's exchange coupling $J$.
The value of ${U=2.56}$\,eV is chosen based on the constrained random-phase approximation (cRPA) analysis conducted in~\cite{PhysRevB.86.165105}.
That Hund's coupling is known to play an important role in the physics of Sr$_2$RuO$_4$~\cite{PhysRevResearch.6.023124, PhysRevB.105.155101, PhysRevB.100.125120}, placing it in the family of Hund's metals~\cite{annurev:/content/journals/10.1146/annurev-conmatphys-020911-125045, georges2024hund}.
We chose ${J=0.35}$\,eV, which provides the most accurate mass enhancement and spin susceptibility in the normal state~\cite{ts6y-zb6m}, while remaining within the range predicted by cRPA~\cite{PhysRevB.86.165105}.\\

\noindent
\textit{Effect of temperature on the spin susceptibility --} In Fig.~\ref{fig:X_BZ_vs_T}, the evolution of the static spin susceptibility with decreasing temperature is plotted on the BZ, for $3\%$ of strain. 
At the second panel, corresponding to ${T=155}$\,K, the secondary peak at $Q'$ is sharper and larger, compared to the first panel (${T=166}$\,K) uncovering the presence of strong competition between the two responses (at $Q$ and $Q'$).
This result is consistent with the incoherent-to-coherent crossover of the $d_{xy}$ orbital at intermediate temperatures, which enhances the magnetic signal at the $Q'$ vector, as discussed in the main text.
The strong interplay between the magnetic fluctuations at the $Q$ and $Q'$ wave vectors is further illustrated in the rightmost panel of Fig.~\ref{fig:X_BZ_vs_T}, which shows the spin susceptibility at $3\%$ strain and a low temperature of ${T=77}$\,K. 
This calculation, performed with a larger chemical-potential shift (${\Delta\mu=0.13}$\,eV), reveals that, for essentially the same electronic filling, an SDW state with ordering vector $Q'$ can also be stabilized.
These findings indicate that the system lies close to a competition between the two SDW instabilities, such that small variations in the electronic filling or numerical accuracy can tip the balance in favor of either the $Q$ or the $Q'$ ordered state.\\

\begin{figure}[b!]
\centering
\includegraphics[width=1\linewidth]{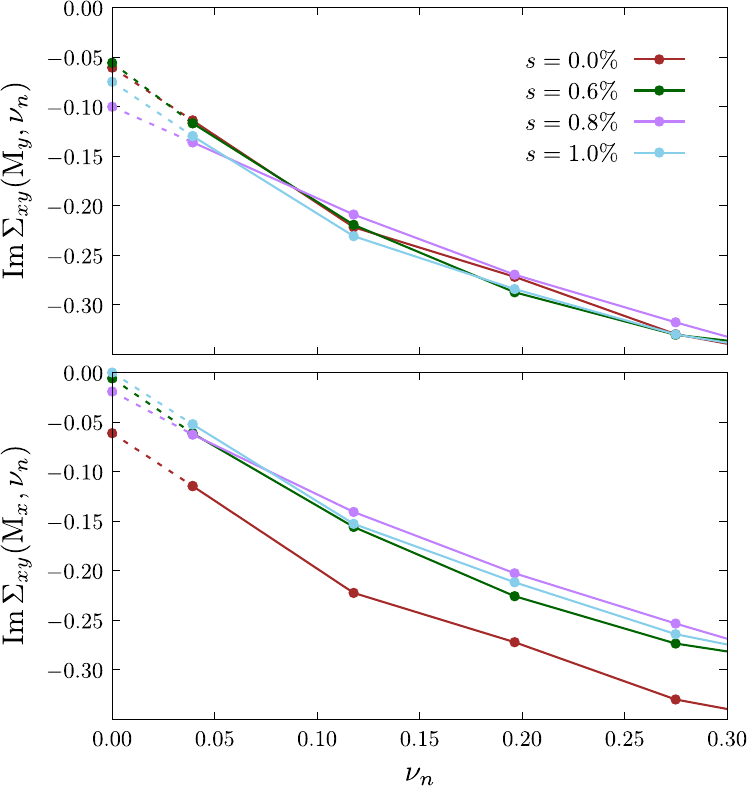}
\caption{\textit{Imaginary part of the self-energies on the Matsubara frequency axis.} Calculations performed for different strain values at ${T=145}$\,K and at two points M$_x$ (lower panel) and M$_y$ (upper panel) of the FS.
\label{fig:S_v_old}}
\end{figure}

\noindent
\textit{Imaginary part of the self-energy on the Matsubara frequency axis --} In Fig.~\ref{fig:S_v_old}, ${\text{Im}\,\Sigma_{xy}}$ is plotted on the Matsubara frequency axis for different values of strain at ${T=145}$\,K, at the M$_x$ and M$_y$ points of the FS. 
As shown in the upper panel, the M$_y$ point retains correlated Fermi-liquid behavior, with ${\text{Im}\,\Sigma_{xy}(\nu\to0)}$ extrapolating to a finite value (scattering rate) of approximately $-0.05$ and $-0.07$ for ${s=0.6\%}$ and ${s=1.0\%}$, respectively.
Importantly, the self-energy exhibits a non-monotonic evolution with strain, with an almost two times enhancement of the scattering rate at $0.8\%$ of strain, namely ${\text{Im}\,\Sigma_{xy}(\nu\to0)\approx -0.10}$. Regarding the M$_x$ point, ${\text{Im}\,\Sigma_{xy}(\nu\to0)}$ gets significantly suppressed compared to the unstrained case, as the system becomes much more coherent and exhibits good FL behavior, with very small scattering rate. The quasiparticle residue Z, obtained through the slope of ${\text{Im}\,\Sigma_{xy}(\nu\to0)}$ remains almost unchanged at the M$_x$ point, while it changes upon strain at the M$_y$ point, with an effect that is however less pronounced than the extrapolated scattering rate.\\

\begin{figure}[t!]
\centering
\includegraphics[width=1\linewidth]{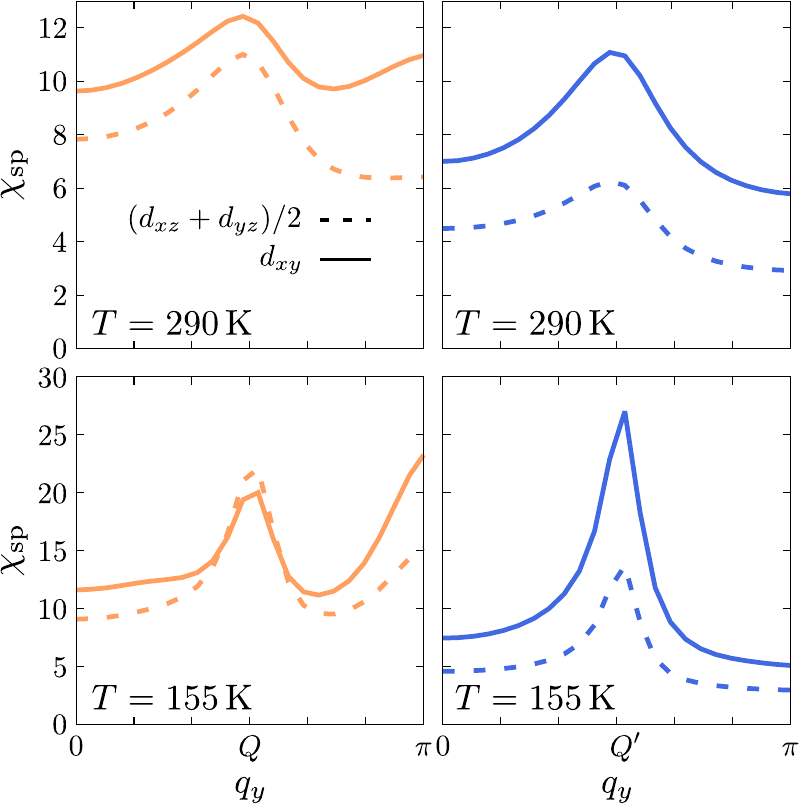}
\caption{\textit{Orbitally-resolved static spin susceptibility.} Calculations are performed on specific paths of the BZ, as mentioned on the third panel of Fig.~\ref{fig:X_BZ_vs_T}, for two temperatures and for $3\%$ strain. 
\label{fig:Xs_Q_Qpr}}
\end{figure}

\noindent
\textit{Orbitally-resolved spin susceptibility --} In Fig.~\ref{fig:Xs_Q_Qpr} the orbitally resolved intra-orbital static spin susceptibility is plotted along the two momentum paths of the BZ mentioned with dashed lines on the third panel of Fig.~\ref{fig:X_BZ_vs_T}, for two values of temperature ${T=290}$\,K (upper panels) and ${T=155}$\,K (lower panels). 
We should mention here that when comparing with Figure~\ref{fig:X_T} of the main text, one should keep in mind that the inverse of the total spin susceptibility (sum of diagonal and off-diagonal terms) is plotted there.
An overall observation is that upon lowering the temperature the response is more sharp, while at higher $T$ the low coherence of the system leads to a heavily broadened magnetic response. 
For both temperatures, the leading contribution of the $d_{xz}+d_{yz}$ orbitals is at the $Q$ vector, as it is linked to their corresponding FS nesting. 
For the higher temperature, the $d_{xy}$ orbital exhibits a slightly higher contribution at $Q$ compared to $Q'$, while upon lowering the temperature the $Q'$ vector response, related to the FS nesting of $d_{xy}$ orbital, is more dominant compared to the $Q$ one.
This result reveals the competing nature of the magnetic response between two vectors linked to the nesting of FS with predominantly different orbital character.

\end{document}